\documentclass[aps,prl,twocolumn,groupedaddress]{revtex4}

\usepackage{graphicx}
\usepackage{amsmath}
\usepackage{amsfonts,amssymb}
\newcommand{\ket}[1]{| #1 \rangle}

\newcommand{\rb}[1]{\left( #1 \right)}

\begin{document}


\title{Quantum Chaos Triggered by Precursors of a 
Quantum Phase Transition: The Dicke Model}


\author{Clive Emary}
\email[]{emary@theory.phy.umist.ac.uk}
\author{Tobias Brandes}
\affiliation{ Department of Physics,
           UMIST,
           P.O. Box 88,
           Manchester 
           M60 1QD, 
	   U. K.}


\date{\today}

\begin{abstract}
We consider the Dicke Hamiltonian, a simple quantum--optical model
which exhibits a zero--temperature quantum phase 
transition. 
We present numerical results demonstrating that  
at this transition the system changes from being 
quasi-integrable to quantum chaotic.  By deriving an exact solution
in the thermodynamic limit we relate this phenomenon to a 
localisation--delocalisation transition in which 
a macroscopic superposition is generated.  We also 
describe the classical analogues of this behaviour.
\end{abstract}

\pacs{05.45.Mt, 42.50.Fx, 73.43.Nq}

\maketitle


At zero temperature, systems of $N$ interacting particles 
can exhibit a quantum phase transition (QPT) as a function 
of a coupling parameter $\lambda$
in the limit that $N\to \infty$. How do the 
precursors of such a transition influence  quantum chaotic (and non-chaotic)
behaviour of the same system for finite $N$?

One of the most direct indicators of the emergence of quantum chaos 
is the change in energy level spacing statistics 
from Poissonian to being described by the Gaussian 
ensembles of Random Matrix Theory.
Although this change-over has been observed in many
systems \cite{guhr,bo:gs,de:ga,ja:sh}, 
only in a comparatively few, isolated cases
has the onset of quantum chaos been correlated with the presence 
of a QPT.  Important 
examples include spin glass shards, which have recently been
used in modeling the onset of chaos in quantum computers \cite{ge:sh}, 
the Lipkin model \cite{LMG}, the interacting boson model \cite{ibm},
and
the three--dimensional Anderson model \cite{le:ra,zh:kr}, where the
change in level statistics occurs at the metal--insulator
(localisation--delocalisation) transition found in disordered 
electronic systems.

In this Letter we consider the Dicke 
Hamiltonian (DH) \cite{di:ck}, a quantum-optical model
describing the interaction of $N$ two-level atoms with a 
number of  bosonic modes.
We demonstrate that a crossover between Poisson
and Wigner--Dyson statistics in this model for finite $N$ is
intimately connected to a mean-field type superradiance QPT.

The simplicity and generality of the Dicke Hamiltonian 
have afforded it appeal both for the investigation of  
quantum chaos, and as a model for phase transitions at a 
critical coupling $\lambda_c$ induced by the interaction with light.   
The level statistics for finite $N$ have
revealed the existence of quantum chaos in certain isolated regimes 
of the model \cite{gr:ho,le:ne}.
On the other hand, the QPT aspect for
$N\to \infty$ has been discussed 
in the context of superradiance \cite{he:li,su:pr},
and recently for exciton condensation \cite{ea:li}.
Here, we derive an {\em exact} solution for all eigenstates, 
eigenvalues and critical exponents in the thermodynamic limit, and show that
above the critical point $\lambda=\lambda_c$ the ground-state 
wavefunction bifurcates
into a macroscopic superposition for any $N < \infty$. 
Our numerical results  indicate that
a localisation-delocalisation transition for 
$N \rightarrow \infty$ underlies
the cross-over between Poissonian and Wigner level--spacing 
distributions for finite $N$.
Furthermore, we use an  exact Holstein-Primakoff 
transformation to derive the classical
limit of the model for arbitrary $N$ and find
a transition at $\lambda=\lambda_c$ from  regular to chaotic trajectories.
The latter are delocalised around {\em two} fixed points in phase space
which we conjecture constitute the classical analogue of the 
macroscopic superposition.
The emergence of this state in the Dicke model is 
related to a conserved parity which becomes spontaneously broken in the 
thermodynamic limit.

We consider the single--mode Dicke Hamiltonian (DH)
\begin{eqnarray}
H = \omega_0 J_z + \omega a^\dagger a 
     + \frac{\lambda}{\sqrt{2j}} \rb{a^\dagger + a}\rb{J_+ + J_-}, 
\label{DHam1}
\end{eqnarray}
where $J_z$, $J_\pm$ are the usual angular momentum operators 
for a pseudo--spin of length $j=N/2$, and $a$, $a^\dagger$ are 
the bosonic operators of the field. The  atomic level--splitting 
is given by $\omega_0$,
$\omega$ is 
the field frequency, and $\lambda$ is the atom--field coupling.  
Crucially, we
have not made the rotating--wave approximation (RWA), as this would 
render the model integrable and destroy
the phenomena that we describe here \cite{le:ne}.
There is a conserved parity $\Pi$  associated with the DH, 
which is given by
$\Pi= \exp\left\{i\pi \left[a^\dagger a + J_z+j \right]\right\}$,
such that $\left[H,\Pi\right]=0$.  The eigenvalues of $\Pi$ 
are $\pm 1$ and, unless stated, we shall work 
exclusively in the positive parity subspace.

We begin by discussing the properties of the 
system in the thermodynamic limit $N,j \rightarrow \infty$.
In this limit the system becomes integrable for all
$\lambda$, and we can derive
effective Hamiltonians to describe the system exactly in each of 
its two phases. We employ a procedure similar to 
Hillery and Mlodinow's analysis of the RWA Hamiltonian \cite{hi:ll}, 
and introduce the
Holstein--Primakoff representation of the
angular momentum operators,
$J_+ = b^\dagger \sqrt{2j - b^\dagger b}$,
$J_- = \sqrt{2j - b^\dagger b}~ b$,
$J_z = \rb{b^\dagger b - j}$,
where $b$ and $b^\dagger$ are bosonic operators \cite{ho:pr}.  Making these 
substitutions allows us to write the DH as a two-mode Hamiltonian.
Below the phase transition, we proceed to the thermodynamic limit
by expanding the square roots and neglecting terms 
with powers of $j$ in the 
denominator. 
This yields the effective Hamiltonian
$
H^{(1)} = \omega_0 b^\dagger b + \omega a^\dagger a
 + \lambda \rb{a^\dagger + a}\rb{b^\dagger + b} - j \omega_0.\label{lcDH1}
$
This bi--linear Hamiltonian may 
then be diagonalised to give 
$H^{(1)} = \varepsilon^{(1)}_- c_-^\dagger c_- 
+ \varepsilon^{(1)}_+ c^\dagger_+ c_+ - j \omega_0$, 
where $\varepsilon^{(1)}_\pm $ are 
the excitation energies of the low--coupling phase, and are
given by 
\begin{eqnarray}
2\rb{\varepsilon^{(1)}_{\pm}}^2 = 
\omega^2 + \omega_0^2 
\pm \sqrt{\rb{\omega_0^2 - \omega^2}^2 
+ 16 \lambda^2 \omega \omega_0}.  
\end{eqnarray}
The energy $\varepsilon^{(1)}_-$ is 
real only for $\lambda \le \sqrt{\omega \omega_0} / 2 = \lambda_c$, 
which locates the phase transition.
We derive an effective Hamiltonian above $\lambda_c$ by first 
displacing each oscillator mode in the Holstein--Primakoff DH by 
a quantity proportional to $\sqrt{j}$, 
and then neglecting terms as above.  With an appropriate choice of 
displacements, this process also yields a bi--linear Hamiltonian, 
which may be diagonalised to a form similar to $H^{(1)}$, but 
with different vacuum and excitation energies,  
the latter of which are given by 
\begin{eqnarray}
2 \lambda_c^4 \rb{\varepsilon^{(2)}_{\pm}}^2 &=&
\omega_0^2\lambda^4 + \omega^2 \lambda_c^4
\nonumber \\ 
&&\pm \sqrt{\rb{\omega_0^2\lambda^4 - \omega^2 \lambda_c^4}^2
+ 4 \omega^2 \omega_0^2 \lambda_c^8}.
\end{eqnarray}
\begin{figure}[t]
\centerline{
  \includegraphics[clip=true,width=1.0\columnwidth]{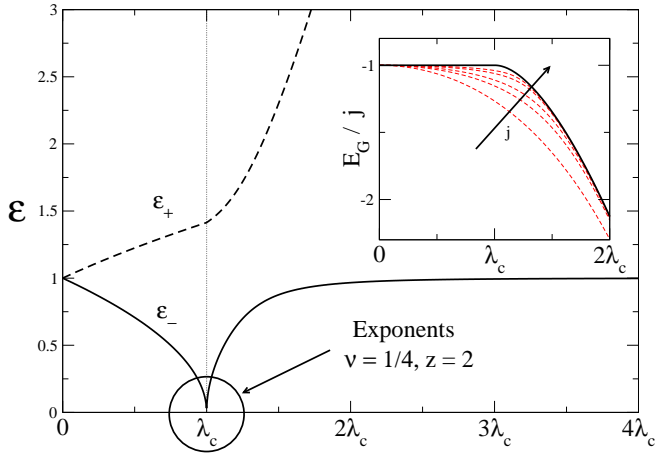}}
\caption{\label{egosc} Excitation energies $\varepsilon_\pm$ 
of the DH in the thermodynamic limit.  Inset:
scaled ground--state energy, $E_G /j$, in the thermodynamic limit 
(solid line) and at various finite values of $j=1/2,1,3/2,3,5$ 
(dashed lines).  The Hamiltonian is 
on scaled resonance $\omega =\omega_0 = 1$.}
\end{figure}
The excitation 
energy  $\varepsilon^{(2)}_-$
of this second Hamiltonian $H^{(2)}$ is real only above the 
phase transition.  There are two independent choices 
of displacements, each of which yield a different effective 
Hamiltonian above $\lambda_c$.  
This is a consequence of the fact that at the QPT the 
$\Pi$ symmetry is spontaneously broken.

In Fig. \ref{egosc} we plot as a function of coupling the behaviour of 
the excitation energies and 
the ground-state energy, which is itself continuous but
possesses a discontinuity in its second derivative  at $\lambda_c$.
Below the phase transition the ground-state is composed of an empty field with 
all the atoms unexcited and hence $\langle J_z\rangle_G/j = -1$ and 
$\langle a^\dagger a\rangle_G/j = 0$.  Above $\lambda_c$, the field is 
macroscopically occupied,
$\langle a^\dagger a\rangle_G/j = 
2\rb{\lambda^4 - \lambda_c^4} / \rb{\omega \lambda}^2$, 
and the atoms acquire a 
macroscopic inversion, 
$\langle J_z\rangle_G/j =-\lambda_c^2 / \lambda^2$.  At the 
phase transition, the excitation energy $\varepsilon_-$ vanishes as 
$ |\lambda-\lambda_c|^{z\nu}$ and
the characteristic length scale $l_- = 1/\sqrt{\varepsilon_-}$ diverges as
$ |\lambda-\lambda_c|^{-\nu}$, with exponents given by $\nu = 1/4$, 
$z=2$ on resonance.

\begin{figure}[t]
\centerline{\includegraphics[clip=true,width=1.0\columnwidth]{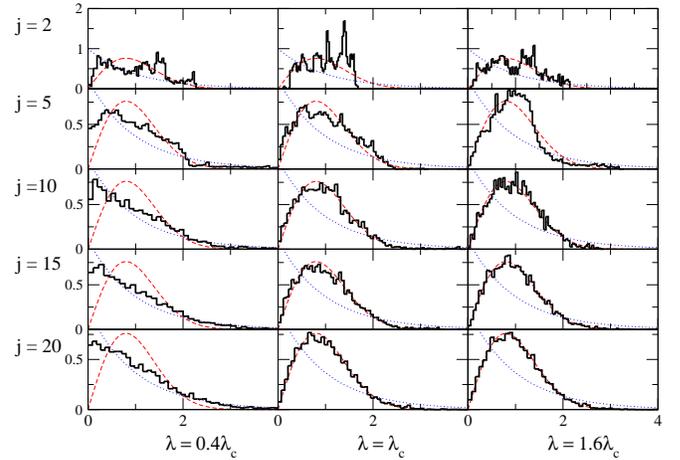}}
\caption{\label{P(S)}Plots of nearest--neighbour distributions $P(S)$ for the 
Dicke Hamiltonian, for different couplings $\lambda$ and pseudo--spin $j$.
Also plotted are the universal Poissonian (dots) to Wigner (dashes) 
distributions.}
\end{figure}
To investigate the level statistics of the system, we 
numerically diagonalise the Hamiltonian in the basis
$\left\{\ket{n}\otimes \ket{j,m}\right\}$, 
where $a^\dagger a \ket{n} = n \ket{n}$, and $\ket{j,m}$ are 
the Dicke states, 
$J_z \ket{j,m} = \rb{b^\dagger b - j}\ket{j,m} = m\ket{j,m}$.
We restrict
ourselves to the positive parity subspace
by considering only states with $n + m + j$ even.
We then unfold the resulting energy spectrum to rid it of 
secular variation,  
form the level spacings  $S_n = E_{n+1} - E_n$, and then construct the
nearest--neighbour distribution function $P\rb{S}$.   Finally, we 
normalise the results for comparison with the universal ensembles 
of Random Matrix Theory.  In the following, we shall use the term 
``quasi-integrable'' to denote systems exhibiting Poissonian 
level statistics, and reserve ``integrable'' for systems
possessing exact solutions.

Figure \ref{P(S)} shows the $P\rb{S}$ distributions obtained for the
DH at various values of the coupling, and for various values of $j$.
At low $j$ the $P\rb{S}$ clearly do not correspond to any of the universal
ensembles.  This is most keenly observed in the $j=1/2$ case 
(identical to the Rabi Hamiltonian \cite{ii:ra}), where the 
spectrum is of ``picket-fence'' character \cite{kus}, characteristic of 
one-dimensional systems or harmonic oscillators \cite{be:ta}.  
For couplings less than
the critical value, $\lambda<\lambda_c$, 
we see that as we increase $j$, the $P\rb{S}$ 
approaches ever closer the Poissonian distribution, 
$P_\mathrm{P}\rb{S} = \exp\rb{-S}$.
At and above $\lambda_c$ 
the spectrum is seen to converge onto the Wigner 
distribution $P_\mathrm{W}\rb{S} = \pi S/2 \exp\rb{-\pi S^2/4}$, 
characteristic of quantum chaos.  This demonstrates that the 
precursors of the QPT in this model lead to a cross--over from 
quasi--integrable to quantum chaotic behaviour 
at $\lambda \approx \lambda_c$.

\begin{figure}[t]
\centerline{
\includegraphics[clip=true,width=1.0\columnwidth]{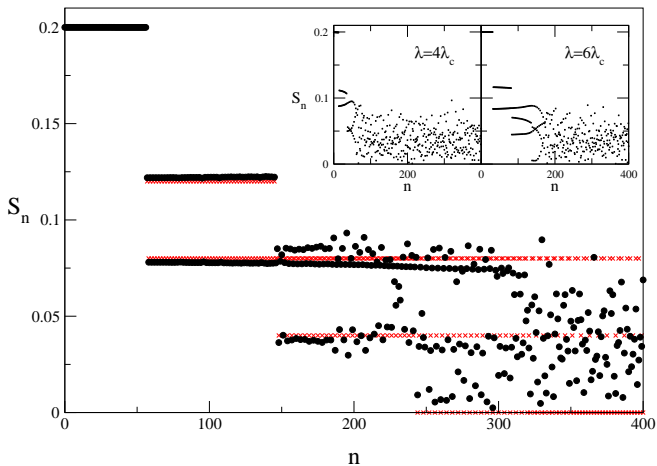}}
\caption{\label{anom}Nearest--neighbour spacing 
$S_n=E_{n+1}-E_n$ vs. eigenvalue 
number $n$ plot for $j=5$ with $\lambda =8\lambda_c$.
Horizontal crosses: results for the integrable 
$\lambda \rightarrow \infty$ Hamiltonian.  Inset: $j=5$ results
with $\lambda =4\lambda_c$ and $\lambda=6\lambda_c$.}
\end{figure}

A further transition between integrable and chaotic
behaviour is observed in the sequence of level spacings $S_n$.  
The $\lambda \rightarrow \infty$ 
limit of the Hamiltonian is integable  for arbitrary  $j$, 
having eigen-energies
$E_{nm} = \frac{\omega}{j} n - \frac{2 \lambda^2}{\omega j^2} m^2$,
where $n=0,1,2,\ldots$ and $m=-j,\ldots,+j$.
As $\lambda$ is increased from $\lambda_c$ to approach this limit
with $j$ fixed, the spectrum reverts from Wigner-like to integrable.  
However, it does not follow the usual transition sequence, but 
rather through a sequence illustrated by Fig. \ref{anom}.
The spectrum 
becomes very regular at low energy, where it approximates the 
integrable $\lambda \rightarrow \infty$ results closely. 
Outside the regular 
region the spectrum is well described by the Wigner surmise, and 
the energy-scale over which the change between the two regimes occurs 
is seen to be surprisingly narrow.  As coupling is increased, the 
size of the low-energy integrable window increases, until it 
eventually engulfs the whole spectrum as $\lambda \rightarrow \infty$.

\begin{figure}[t]
\includegraphics[clip=true,width=0.71\columnwidth]{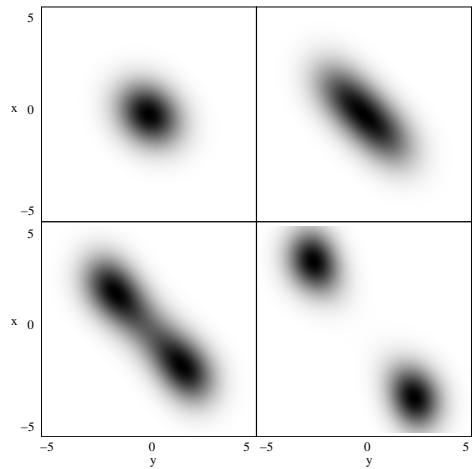}
\caption{\label{finj} The modulus of the ground-state wavefunction 
$\psi\rb{x,y}$
of the Dicke Hamiltonian in the abstract $x$-$y$ representation  
for finite $j=5$, at couplings of 
$\lambda / \lambda_c$ = 0.4, 1.0, 1.2, 1.4.  Black corresponds 
to $\mathrm{Max}|\psi|$ and white corresponds to zero.}
\end{figure}

We now proceed to consider the wavefunctions of the system by
introducing an abstract position--momentum representation 
for each of the boson modes via
$x \equiv \frac{1}{\sqrt{2 \omega}}\rb{a^\dagger + a}$; 
$p_x \equiv i\sqrt{\frac{\omega}{2}}\rb{a^\dagger - a}$, and 
$y \equiv\frac{1}{\sqrt{2 \omega_0}}\rb{b^\dagger + b}$; 
$p_y \equiv i\sqrt{\frac{\omega_0}{2}}\rb{b^\dagger - b}$.
In this representation the action of the parity operator $\Pi$ 
corresponds to rotation by $\pi $ about the origin.
The ground state of the system on 
scaled resonance for $j=5$ is plotted
for different couplings in Fig. \ref{finj}.  
These wavefunctions
were obtained by diagonalising the Hamiltonian in the same basis as was
used to calculate $P\rb{S}$ and representing the basis vectors
$\ket{n}\otimes\ket{m,j}$ by products of harmonic oscillator 
eigenfunctions.
For the non-interacting system ($\lambda=0$), the wavefunction 
is a product of two independent Gaussians.  As the coupling increases,
the two modes start mixing, leading to a stretching of the 
single--peaked wavefunction.  Around 
the critical coupling ($\lambda \approx \lambda_c$), the 
wavefunction bifurcates to become a double--peaked function, 
and with further increases in coupling the two lobes 
at $\rb{\pm x_0, \mp y_0}$
move away from each other in their respective quadrants of the $x-y$ plane.
Since $x_0$ and $y_0$  are of the order of $\sqrt{j}$, for 
large $j$ the ground state evolves into a superposition of two 
macroscopically distinguishable parts, which may be
considered as a ``Schr\"{o}dinger's cat''.
The formation of this state constitutes a delocalisation of the ground-state
wavefunction, which is also observed in the excited states.
This localisation-delocalisation transition is consistent with the
transition between Poisson and Wigner distributions in the spectrum 
\cite{prange}.
The suppression of chaos at low $j$ is then seen to be due to 
the fact that for 
low $j$ only a few excitations are permitted in the $b$-mode.
This restricts the extent of the wavefunction in the $y$-direction, 
inhibiting delocalisation, and yielding the non-generic $P\rb{S}$ seen in 
Fig. \ref{P(S)}.
It should be noted that 
an actual spontaneous symmetry--breaking transition occurs
above $\lambda_c$ in the $j \rightarrow \infty$ limit which removes the two 
lobes so far from one another that
they cease to overlap and thus can be considered independently. 
This breaks the $\Pi$ symmetry of the model, allowing us to obtain 
the earlier exact results by using an effective 
Hamiltonian for each lobe.  These Hamiltonians are identical in form,
have identical spectra, thus
demonstrating that in the thermodynamic limit, each energy level in the  
high--coupling phase is doubly degenerate and 
the macroscopic superposition is broken in half.


Finally, this position representation allows us
to study the classical analogues of this QPT and the accompanying 
onset of chaos in a very natural way.   By setting
the commutators $\left[x,p_x \right]$ 
and $ \left[y,p_y \right]$ to zero, the classical
Hamiltonian corresponding to Eq. (\ref{DHam1}) is seen to be
\begin{eqnarray}
H_{\mathrm{cl}} &=&
-j \omega_0 + \frac{1}{2}
\rb{
  \omega_0^2 y^2 + p_y^2 + \omega^2 x^2 + p_x^2 
  -\omega -\omega_0} \nonumber \\
&&+ 2\lambda \sqrt{\omega\omega_0}x y 
\sqrt{
  1 - \frac{1}{4\omega_0 j}\rb{\omega_0^2 y^2 + p_y^2  
  - \omega_0}
}\label{clDH}.
\end{eqnarray}
\begin{figure}[tb]
\centerline{
\includegraphics[clip=true,width=1.0\columnwidth]{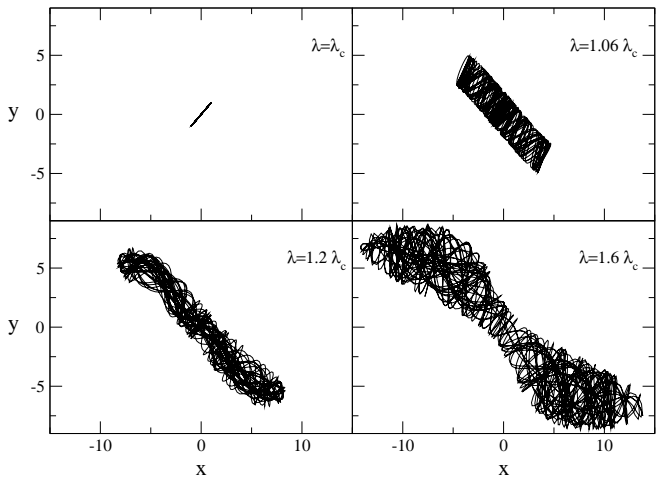}}
\caption{\label{clas}  Typical classical phase space projections
($p_x=p_y =0$) for the classical Dicke Hamiltonian of Eq. (\ref{clDH})
with $\lambda / \lambda_c$ = 1.0, 1.06, 1.2, 1.6, for 
$j=30.0$, $\omega=\omega_0=1$.
Initial conditions were $x\rb{0}=y\rb{0}=1$, $p_x\rb{0}=p_y\rb{0} =0$.
The abrupt
change to complex motion is observed above  $\lambda=\lambda_c$}
\end{figure}
Space limitations here only allow us to point out two significant 
features. 
Firstly, below $\lambda_c$ there is only one fixed point of the flow, namely
$x=y=p_x=p_y=0$.  At $\lambda = \lambda_c$ this situation changes abruptly and
two new fixed points appear at $\rb{x,y}=\rb{\pm x_{cl}, \mp y_{cl}}$  
with $p_x =p_y=0$,
where $x_{cl}$ and $y_{cl}$ are approximately equal to the centres
of the wavefunction lobes $x_0, y_0$ for large $j$. 
Secondly, parametric plots
of typical trajectories obtained from Eq. (\ref{clDH}) (Fig. \ref{clas})  
demonstrate that the system undergoes 
a rapid change at $\lambda = \lambda_c$ from a very simple 
quasi-periodic motion to intricate chaotic behaviour, 
in agreement with the results of the quantum model.
Note that the correspondence between this classical system 
and the original quantum one is significantly 
greater than previous semi-classical treatments \cite{mil}, 
and that this holds for any $j$, not just for $j \rightarrow \infty$.


We mention that larger system sizes are required to check if there 
exists a critical level statistics of our model 
at $\lambda_c$ \cite{zh:kr}, and that an examination of the 
exceptional points \cite{LMG} of this model may yield further insight.
Future work also includes a study of related models to
determine the generality of the features described here.

In summary, we have seen that the $P\rb{S}$ distribution of the DH
at finite $N$ changes from being Poissonian to Wigner at approximately
$\lambda_c$, indicating the emergence of quantum chaos.  
The ground-state wavefunction bifurcates at this point, 
forming a macroscopic superposition.
The underlying quantum phase transition is reflected 
in the classical model derived here by the appearance of two new 
fixed points at $\lambda_c$, where a transition between 
regular and chaotic trajectories occurs.

\begin{acknowledgments}
This work was supported by projects EPSRC GR44690/01, DFG Br1528/4-1,
the WE Heraeus foundation and the UK Quantum Circuits Network. 
\end{acknowledgments}

\end{document}